# DC-Augmented Dielectric Barrier Discharge (DCA-DBD)

A. Tang[1], A. Aliseda[1], A. Mamishev[2], I. Novosselov[1*]

[1] *Mechanical Engineering Department, University of Washington, Seattle, WA*
[2] *Electrical and Computer Engineering Department, University of Washington, Seattle, WA*
*\* Corresponding Author: ivn@uw.edu*

*Abstract*— **Time-dependent multiphysics interactions that drive the energy transfer in electromechanical systems are poorly understood. We probe dielectric barrier discharge (DBD) with an external DC-augmented (DCA) field to reveal new mechanistic insights. The biased HV DC electrode influences the interaction between the charged ions and the E-field, surface and space charge, and neutral molecules. Direct force measurement, velocity profiles, and time-resolved electrical and optical measurements of discharge characteristics provide evidence of complex plasma/flow interactions. Negative DCA leads to modest improvements in momentum transfer due to the field-augmented ion acceleration before the system transitions to sliding discharge and a counter jet at the DCA electrode, canceling the gains from positive ion acceleration. Positive DCA monotonically increases the wall-parallel force. A new oscillating residual charge interaction mechanism is identified to explain a greater than 2-fold increase in horizontal thrust, in which the acceleration of positive ions is augmented by the attraction from the residual (negative) charge.**

Non-thermal plasma devices have received significant interest from the scientific and engineering perspectives [1-3]. Plasma discharge has been utilized in surface treatments [4-6], particle matter collectors [7-9], and mass spectrometry applications [10]. Momentum injection by plasma devices has been used for propulsion [11-14] and active flow control [15-17]. Non-thermal plasmas such as corona discharge or dielectric barrier discharge (DBD) generate a flow by first ionizing a fraction of the gas molecules and then accelerating free electrons and ions in the electric field. An electrohydrodynamic (EHD) force is generated through collisions with air molecules. In corona discharge, the ionization occurs between two high voltage (HV) direct current (DC) electrodes [18, 19], while DBD actuators use pulsed or alternating current (AC) [20, 21].

A single-stage DBD actuator consists of an air-exposed electrode, an embedded electrode, and a dielectric barrier between the two. Scientific literature has characterized EHD wall jets for two-electrode DBDs [22-24]. Currently, their applications for flow controls and propulsion are limited due to modest forcing, motivating the need to study multi-electrode DBD actuators [25-27]. However, the mechanistic understanding of the complex coupling between electric characteristics, surface charging, recombination reactions, and charge–flow interaction is still speculative, partly due to the large time-scale separation.

In AC-driven DBD, the discharge–plasma–flow interaction can be divided into positive and negative voltage half-cycles. One of the most accepted mechanisms is 'push-push' forcing, where positive ions are repelled from the active electrode by Coulombic forces during the positive half-cycle, and negative ions are repelled during the negative half-cycle [26]. In actuality, the physics is more complicated. Electric current measurements and high-speed plasma visualizations show that the positive half-cycle leads to streamer-like discharge, and the negative half-cycle produces smaller and more frequent discharges associated with glow discharge [28, 29]. Though the discharge current is higher in the positive half-cycle [22], some reports suggest that the forcing terms can be as strong in the negative half-cycle due to the high mobility of electrons [23, 30]. Unlike in corona discharge, where the steady-state forcing term can be derived analytically [11, 18], elucidating the time-dependent plasma–flow interaction in DBD is challenging. Insights into the time-resolved behavior can be used to enhance the energy transfer and minimize parasitic losses.

DBD forcing on a fluid volume can be increased by introducing a third electrode [31, 32]. In this arrangement, the DBD electrode pair ionizes the gas, and the biased third electrode accelerates positive or negative species, promoting their interaction with neutral molecules. This scheme can be named DC-augmented DBD (DCA-DBD).

This work investigates DCA-DBD with an asymmetric AC electrode pair and a third HV DCA electrode, which is biased with either polarity to preferentially interact with the positive or negative half-cycle, serving as a probe into the time-dependent interaction. Since the coupling of the AC E-field, space and surface charges, and fluid dynamics is highly non-linear. The DCA probes several aspects of the system, and understanding these effects requires multiple lines of evidence.

The experimental setup (Figure 1) consists of the primary DBD electrode pair, separated by the dielectric layer, a 3.15 mm quartz plate with 99.995% $SiO_2$ purity. The length of the air-exposed electrode is 15 mm, and the encapsulated electrode is 25 mm. The edges of the electrodes are aligned. The encapsulated electrode is covered by thick polyimide and silicone rubber tape (~6.3 mm) to prevent backside discharge. The third air-exposed electrode is 50 mm downstream of the first electrode. Thrust, velocity, and current measurements are normalized by the actuator's spanwise width of 110 mm.

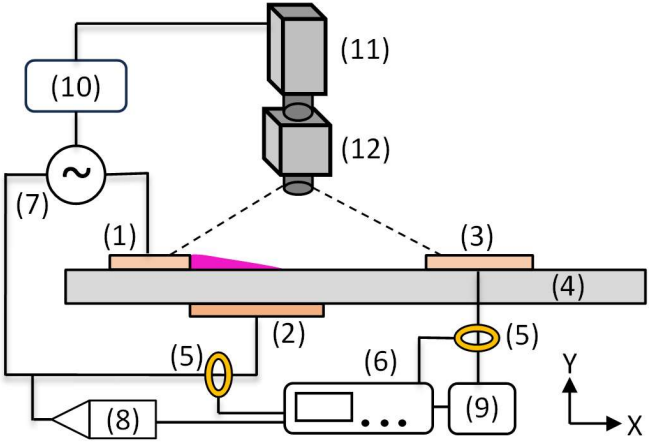

Figure 1. The experimental schematic of the three-electrode DBD actuator comprised an air-exposed electrode (1), embedded electrode (2), air-exposed third electrode (3), and dielectric layer (4). The current through the DBD electrode pair and the third electrode is measured with a non-intrusive current monitor (5) and oscilloscope (6). The electrodes are powered by a custom power supply (7) with a high-voltage probe (8) and a high-voltage amplifier (9). The custom power supply is controlled by a multi-channel function generator (10). The function generators control the trigger outputs to the CCD camera (11) and the UV intensifier (12).

The DBD electrodes are connected to a custom HV power supply comprised of a function generator (Siglent SDG1032X) and a power amplifier (Crown XLi3500) that drives two custom transformers (Corona Magnetics, Inc). The AC frequency is set at $f = 2$ kHz. The voltage output is monitored using a Tektronix 6015A HV probe. The DCA electrode is powered by a Trek 40/15 HV amplifier; the $V_{DCA}$ voltage was varied in the $-24$ to $+24$ kV range.

The mechanical performance was characterized by wall jet velocity profiles and direct thrust measurements. A custom glass pitot tube (OD $=0.6$ mm, ID $= 0.4$ mm) measured x-velocity profiles using an Ashcroft CXLdp differential pressure transmitter ($P = 0 - \pm25$ Pa, 0.25% accuracy) reconstructing the time-average velocity field [33]. The non-conductive pitot tube allowed measurements close to the plasma region while avoiding electromagnetic interference (EMI. To avoid interference, velocity measurements were not taken inside the plasma region. The typical velocity measurements had a standard deviation of $< 0.02$ ms$^{-1}$ when averaged over 20 s. The thrust was measured directly using a battery-powered, electrically insulated analytical balance (Ohaus SPX223) with 1 μN sensitivity. The balance was placed in a Faraday cage to avoid EMI. Both horizontal and wall-normal thrusts were recorded. The balance measurements were transferred to a computer via an optically isolated cable and averaged over 15 seconds.

Figure 2 shows that for all $V_{DCA}$ conditions, the thrust increases with applied AC voltage, which generates more ions and stronger electrical fields. These effects are well documented in the DBD systems [16, 26]. The increase in the negative DCA potential had only a moderate trust increase at the low DCA voltages, followed by a significant drop at

$V_{DCA} \sim -15$ kV. The slight increase is likely due to the *field-augmented species acceleration (FASA)* enabled by the stronger E-field from negatively biased DCA. The drop in the streamwise thrust corresponds to the increase in wall-normal force, suggesting that the EHD wall jet is deflected from the streamwise to the wall-normal direction (Figure 2b). Positive DCA bias leads to a monotonic increase in the EHD thrust. At the highest tested AC potential condition ($V_{AC}$=40 kV, $V_{DCA}$=+24 kV), the horizontal thrust increased by $\sim$50% (45 mN/m to 67 mN/m). In the less energetic ionization case, $V_{AC} = 25$ kV, the thrust increases by 133% (8.2 mN/m to 19.1 mN/m). Explaining these gains by the field-augmented negative species acceleration, independently of the rest of the physics, is difficult. There is a strong possibility of an additional complementary mechanism, which we'll discuss after presenting optical and electric measurements.

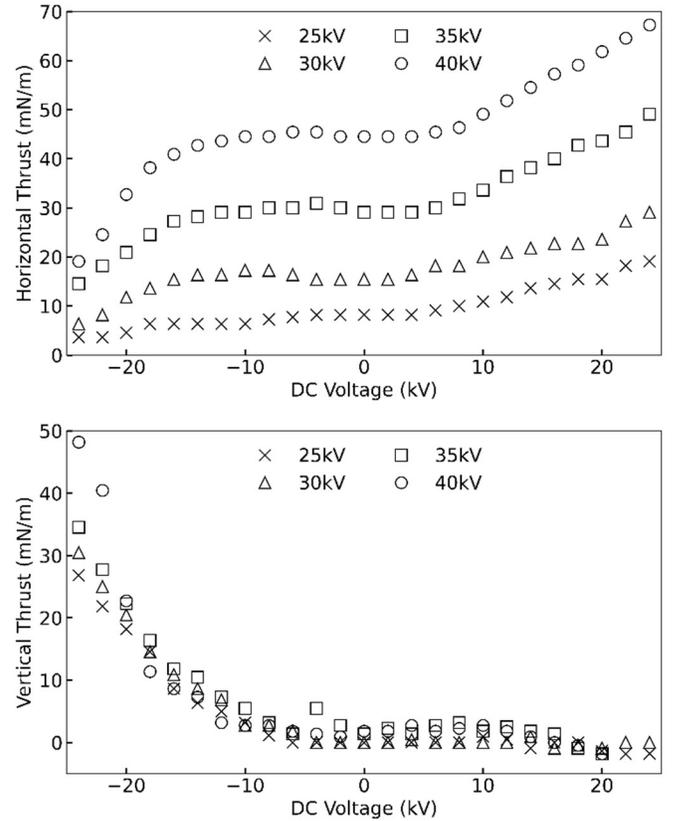

Figure 2. (a) Horizontal and (b) vertical thrust of the DBD-DCA actuator, $f = 2$ kHz, $AC$ voltage is varied 25 kV $< V_{AC} <$ 40 kV; DCA voltage $-24$ kV $< V_{DCA} <$ +24 kV.

The reconstructed 2D velocity contours in Figure 3 show the EHD jet at $V_{AC} = 35$ kV and $V_{DCA} = 0$ kV, $-24$ kV, and +24 kV. The velocity flowfield was constructed by spanning the x-velocity measurements in the region between the air-exposed electrodes, with a 1 mm spacing in the x- and y-directions.

Figure 3b shows that a reverse EHD jet is formed at the third electrode in the negative DCA case. Colliding the primary and reverse jets results in the wall-normal flow and the vertical reaction force (Figure 2b). These wall-normal synthetic jets have been reported in the planar symmetric 3-electrode systems and axisymmetric DBD actuators [34, 35]. Integration of the velocity profiles at x=15 mm, before deflection, shows a 28% increase in x-momentum over the unbiased case. This supports the observation that a slight increase of the horizontal thrust at the low values of negative DCA is due to field-augmented (positive) ion acceleration (Figure 2a).

Consider the positive DCA scenario; for $V_{DCA}$=+24 kV, the maximum velocity, $V_{max}$=5.3 m/s, is slightly higher than that of the unbiased case, $V_{max}$=5.0 m/s. However, the direct thrust measurements show significantly higher thrust (+68%), which is consistent with a higher wall jet thickness (SI Figure 1). In general, the velocity-derived thrust from profile integration closely agrees with the direct thrust measurements (SI Table 1). These results suggest that one of the factors responsible for stronger forcing is the repulsion of positive ions and the EHD jet by the positive DCA, leading to lower wall shear stress. However, the gains in the thrust could not be attributed to only viscous effects, which typically account for <30% reduction in EHD wall jet momentum [11, 36]. The enhancement must also come from the additional momentum transfer mechanism between the ions and the gas. Moreau *et al.* [34] suggested that there could be a significant contribution in forcing due to an acceleration of negative discharge during the negative half-cycle despite the repulsion of the positive discharge, or what we termed here a *field-augmented species acceleration*.

We propose that the effectiveness of the positive DCA can be linked to the extended interaction of positive ions and the space and surface charges. The primary difference in the velocity profiles between unbiased and positive DCA is the jet thickness and the location of the velocity peak. Interactions of positive ions with neutral molecules are primarily localized to the visible plasma region [37, 38]. Thus, as expected, the peak velocity is located at the edge of the visible plasma region, e.g., x= 10 mm in Figure 3a and b. In contrast, in positive DCA (Figure 3c), the flow continues to accelerate with the maximum velocity location at x=12 - 15 mm, indicating that there is an additional pull on the positive ions that could come from the negative space or surface charge generated in a previous (negative) half-cycle. In this "bootstrapping" scenario, the positive DCA bias forces high-mobility electrons to travel and be retained closer to the third electrode during the negative-going half-cycle, providing an additional attraction for positive ions in the following half-cycle. This behavior repeats each AC cycle and can be named the *oscillating residual charge interaction (ORCI)* mechanism.

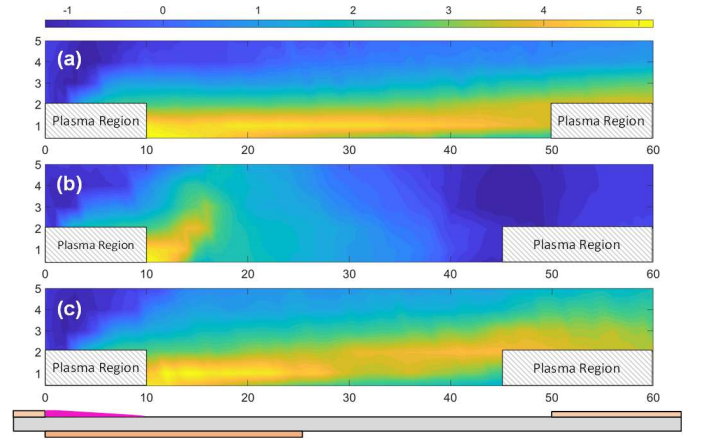

Figure 3. X-velocity contour plot at $V_{AC}$ = 35 kV with (a) $V_{DCA}$ = 0 kV, (b) $V_{DCA}$ = −24 kV, and (c) $V_{DCA}$ = +24 kV. Grayed-out regions had visible plasma discharge; the data in these regions were not taken. The electrode positions are displayed.

Examination of electrical and optical characterization of the discharge provides additional insight into the *FASA* and *ORCI* mechanisms. Figure 4 shows the time-resolved DBD and the DCA electrode currents measured using a non-intrusive high-bandwidth coil current monitor (Pearson 2877 with 2 ns rise time). The monitor is connected to a Tektronix DPO7254C oscilloscope (500 MHz bandwidth). At least 20 voltage cycles were recorded at 1 GS/s, and the total discharge current was obtained by filtering out the capacitive current using a fast-Fourier transform (FFT) analysis [22]. The standard deviation of discharge current measurements for 20 consecutive cycles was < 4% for all tested conditions.

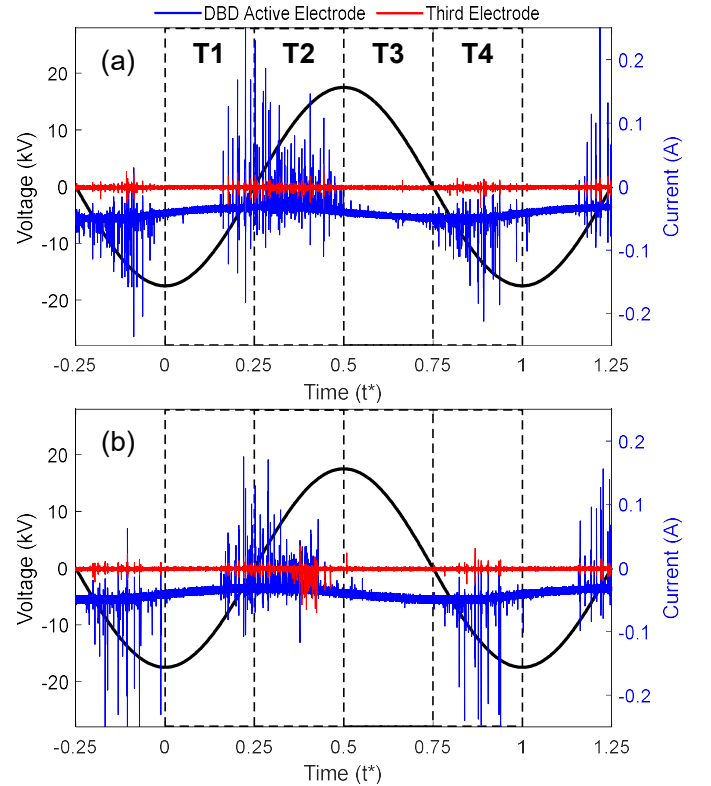

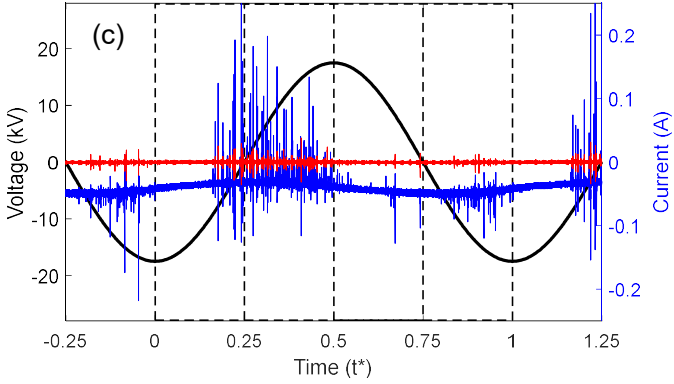

Figure 4. Trigger frames (dashed lines) superimposed with current on DBD active electrode (blue) and DCA electrode (red), $V_{AC} = 35$ kV, $f = 2$kHz; (a) $V_{DCA} = 0$ kV, (b) $V_{DCA} = -24$ kV, and (c) $V_{DCA} = +24$ kV. The $t^*$ is the period normalized time.

Figure 5 shows phase-resolved images of optical plasma emissions from a high-speed CCD camera (Phantom V12.1) equipped with a 200 – 550 nm UV intensifier lens (Specialized Imaging SIL3). The DBD discharge in atmospheric air produces light emissions from positively charged nitrogen and oxygen ions in the wavelength range of 300 – 450 nm [37, 38]. The CCD shutter is synchronized with the voltage cycle, divided into four quadrants denoted as T1 – T4 (Figure **4**). For $f = 2$ kHz, the CCD and intensifier trigger were set to 8 kHz with a CCD shutter of 124 μs and intensifier shutter of 110 μs.

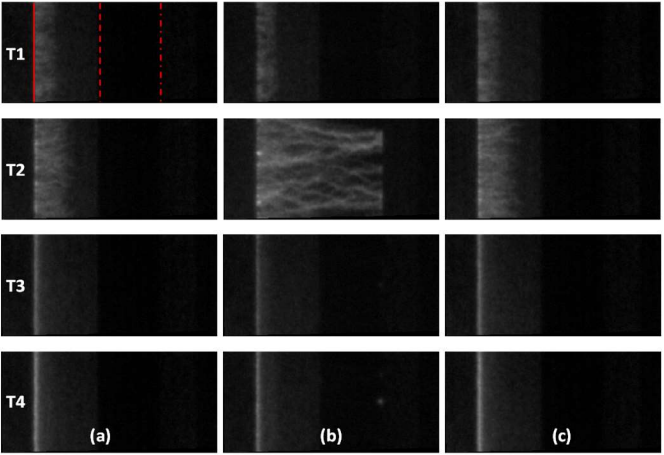

Figure 5. Phase-resolved plasma discharge images with DCA electrode; $V_{AC} = 35$ kV, $f = 2$ kHz (a) $V_{DCA} = 0$, (b) $V_{DCA} = -24$ kV, (c) $V_{DCA} = +24$ kV. T1 and T2 are positive-going cycles, showing a high concentration of positive ions; T3 and T4 are negative-going cycles, with low optical emission at the edge of the active electrode. The leading edge (**–**), embedded electrode edge (**– –**), and third electrode edge (**–.–**) are shown in the T1 phase of (a).

The positive discharge current, $I_{discharge}^+$, in the T1 and T2 regions corresponds to electron emissions during the production of positive ions. This discharge mode is associated with the formation of streamers [29] (Figure 5). In the T3 and T4 regions, the negative discharge current ($I_{discharge}^-$) is associated with electrons leaving the air-exposed electrode, creating a glow discharge. The discharge current in the positive-half-cycle is typically higher, e.g., at $V_{AC} = 40$kV, the $I_{discharge}^+ = 11.2$ mA/m vs. $I_{discharge}^- = 3.65$ mA/m, resulting in the overall abundance of positive charges in the system. A biased $V_{DCA}$ electrode changes the electrical field strength, influencing the positive discharge characteristics in the primary ionization zone. For example, for the $V_{AC} = 40$ kV, $V_{DCA} = +24$ kV case, the positive discharge current is suppressed by 9.4% and is 9.26% greater in the negative DCA case, $V_{DCA} = -24$ kV. This is analogous to the corona discharge, where the higher field strength leads to higher plasma volume and charge density [18, 19]. Interestingly, the $I_{discharge}^-$ in the T3 and T4 quadrants is unaffected by positive and negative $V_{DCA}$ (0.02 mA/m change). It is likely that negative species rapidly recombine with the abundant positive charges in the system.

Under high negative bias ($V_{DCA} < -15$kV), the visible plasma discharge extends to the DCA electrode, resulting in the 'sliding' DBD (Figure 5b). The high concentration of positive ions in the sliding discharge is shown in SI Figure 2. The increasing field strength in the positive half-cycle shifts the distribution of positive surface charge towards the negatively biased DCA electrode. At sufficiently high E-field, the combination of positive surface and space charge causes a negative discharge on the DCA electrode, Figure 4b at $t^*\sim0.4$ in the T2 region. Time-resolved plasma emission images show that the sliding positive discharge streamers are present only in the T2 region of the cycle, i.e., during the strongest E-field. Integration of the discharge peaks on the DCA electrode (Figure 4) yields $I_{discharge, DCA}^- = 3.1$ mA/m, indicating that electrons leave the DCA electrode to recombine with the positive streamers. The electrons' motion creates a counter-flow EHD jet (Figure 3b), which produces wall-normal thrust upon collision with the primary DBD jet, Figure 2 (b). This negative DCA discharge to the positive space charge is analogous to a pulsed corona discharge or a DBD to the space charge. In weaker negative potential DCA cases, the space charge is not sufficiently extended to trigger DCA discharge. Minor horizontal thrust improvements were observed due to the positive ion FASA mechanism (Figure 2a).

For all tested positive DCA conditions, we observed a steady increase in horizontal thrust; two complementary mechanisms can be responsible. (1) *Field-augmented (negative) species acceleration* is enabled by the stronger E-field. Benard *et al.* [30] suggested that the momentum transfer of negative species during the negative half-cycle constitutes most of the wall jet forcing. Without time-resolved force and velocity measurements, we could not evaluate the relative contribution of the FASA mechanism. (2) *ORCI mechanism*, in which the positive DCA enables propagation of negative species towards the third electrode and their retention as a surface or space charge. The negative residual surface charge then attracts the positive ions further downstream in the sequential discharge cycle, extending their interaction with neutral molecules in a streamwise direction. Figure 5 (c, T2) and SI Figure 2 show that the visible discharge zone extends further downstream in positive DCA cases. In some way, this mechanism is similar to ambipolar diffusion, where space charge interacts with the species of opposite polarity [39, 40].

In conclusion, this letter is the first report on the mechanisms of a DC-augmented DBD actuator. A slight increase in the thrust values for the moderate negative DCA due to the field-augmented (positive) ion acceleration and a sufficiently strong field leads to the onset of the reverse flow at the DCA electrode, producing a deflected jet with wall-normal forcing. A positive DCA monotonically improves horizontal thrust > 100%. A field-augmented acceleration and new oscillating residual charge interaction mechanisms are proposed. The contribution of these mechanisms needs to be confirmed and their relative weight further evaluated. Future work should focus on understanding the spatiotemporal distribution of surface and space charges, essential in gaining insights into the DCA-DBD system.

**Acknowledgments**: We thank Dr. Michael Barbour for helping with the Phantom high-speed CCD camera.

**Conflict of Interest**: The authors have no conflicts to disclose.

**Author Contributions:**

**Anthony Tang:** Investigation and conceptualization (equal), Data Acquisition, Writing – original draft. **Alberto Aliseda:** Writing – review and editing (equal). **Alexander Mamishev:** Writing – review and editing (equal). **Igor Novosselov:** Supervision, Writing – review and editing (equal), Investigation and conceptualization (equal).

**Data Availability:**

The data that supports the findings of this study are available from the corresponding author upon reasonable request.